\documentstyle[a4wide,12pt,epsf]{article}
\begin{document}
\begin{flushright}
\baselineskip=12pt
{SUSX-TH-00-010}\\
{RHCPP 00-02T}\\
{hep-ph/0007206}\\
{July  2000}
\end{flushright}

\begin{center}
{\LARGE \bf Sparticle spectrum and dark matter in type I string theory 
with an intermediate scale \\}
\vglue 0.35cm
{D.BAILIN$^{\clubsuit}$ \footnote
{D.Bailin@sussex.ac.uk}, G. V. KRANIOTIS$^{\spadesuit}$ \footnote
 {G.Kraniotis@rhbnc.ac.uk} and A. LOVE$^{\spadesuit}$ \\}
	{$\clubsuit$ \it  Centre for Theoretical Physics, \\}
{\it University of Sussex,\\}
{\it Brighton BN1 9QJ, U.K. \\}
{$\spadesuit$ \it  Centre for Particle Physics , \\}
{\it Royal Holloway and Bedford New College, \\}
{\it  University of London,Egham, \\}
{\it Surrey TW20-0EX, U.K. \\}
\baselineskip=12pt

\vglue 0.25cm
ABSTRACT
\end{center}

{\rightskip=3pc
\leftskip=3pc
\noindent
\baselineskip=20pt
The supersymmetric particle spectrum is calculated in type I string 
theories formulated as orientifold compactifications of type IIB 
string theory. A string scale at an intermediate value of $10^{11}-10^{12}$
GeV is assumed and extra vector-like matter states are introduced to 
allow unification of gauge coupling constants to occur at this scale. 
The qualitative features of the spectrum are compared with Calabi-Yau 
compactification of the weakly coupled heterotic string and with the 
eleven dimensional supergravity limit of $M$-theory. Some striking 
differences are observed. Assuming that the lightest neutralino provides 
the dark matter in the universe, further constraints on the sparticle 
spectrum are obtained. Direct detection rates for dark matter are estimated. }

\vfill\eject
\setcounter{page}{1}
\pagestyle{plain}
\baselineskip=14pt

In a generic supergravity theory, the soft supersymmetry-breaking terms 
are free parameters. On the other hand, if the supergravity theory is the 
low energy limit of a string theory, these parameters are calculable in 
principle in terms of the fewer parameters characteristic of string 
theory. Once the soft supersymmetry-breaking terms have been determined the 
renormalization group equations may be run from the 
string scale to the electroweak scale 
to derive the sparticle spectrum. Such calculations have been performed 
in the context of the weakly coupled heterotic string in the large 
modulus limit of Calabi-Yau compactifications
\cite{IBA:Spain}, of orbifold compactifications 
of the weakly coupled heterotic string \cite{IBA:Spain,ALEX} and of the eleven 
dimensional supergravity limit of M-theory corresponding to the 
strongly coupled heterotic string \cite{BKL,MUNOZ}.
Here we extend such calculations to scenarios motivated by type I string 
theories constructed as orientifold compactifications of type IIB string 
theory \cite{FERNANDO,URANGA}. A novel feature of 
type I theories is that the string scale is a 
function of the Planck scale and the compactification scale and can, in 
principle, lie anywhere between about 1TeV and $10^{18}$GeV \cite{witten}-
\cite{Karim}. A rather 
natural possibility is for the string scale to be at an intermediate scale 
of order $10^{11}$GeV. Type I theories possess an elegant mechanism for this 
to occur which may be summarized as follows. It is possible to construct 
type I theories in which the observable gauge group and quark and lepton matter are associated with 9-branes and 5-branes while supersymmetry is broken 
directly in non-supersymmetric anti-5-brane sectors. The scale of 
supersymmetry breaking in the anti-5-brane sector is the type I string 
scale $M_I$ and the supersymmetry breaking will be transmitted gravitationally 
to the observable sector. We then expect masses for the supersymmetric 
particles of order $M_I^2/M_P$ where $M_P$ is the Planck mass. For sparticle 
masses of order $1$TeV we have $M_I\sim 10^{11}$GeV. In a string theory, 
unification of (tree-level) gauge coupling constants will occur at $M_I$.

Our purpose here is to study the consequences of such an intermediate 
string scale for the sparticle spectrum, and for dark matter assuming 
that the lightest neutralino provides the dark matter in the universe.
A novel sparticle spectrum can develop because the renormalization group 
equations are being run between the electroweak scale and an intermediate 
scale rather than a scale of order $10^{16}$GeV and because unification 
of gauge coupling constants at $M_I$ may require extra matter (vector-like) 
between the electroweak scale and $M_I$. There is some overlap with the 
work of Abel et al. \cite
{IBANEZ} which appeared in e-print when our work was close 
to completion. However, we consider some choices of extra matter to achieve 
intermediate scale unification and some choices of Goldstino angle which differ from these authors and also explore the possibility of an unconventional 
normalization of the $U(1)$ of the standard model, which is common in 
type I theories. We also study the implications of the sparticle spectrum 
in type I theories for dark matter. There is also some overlap with the 
work of Gabrielli et al \cite{BOTT} which appeared in e-print while 
we were writting up this work. These authors also considered direct detection 
rates for neutralino dark matter in theories with intermediate scales.

We shall study two scenarios which illustrate how far the type I sparticle 
spectrum can differ from the sparticle spectrum obtained in Calabi-Yau 
compactifications of the weakly coupled heterotic string and from the 
eleven dimensional supergravity limit of M-theory corresponding to the 
strongly coupled heterotic string.
In the first scenario ($I_a$), the additional matter, taken to have mass 
small on the intermediate scale but large compared to 
``observable'' matter, is taken to be $2L+3E_R$ vector-like representations, 
where $L$ is an $SU_L(2)$ lepton-like 
doublet and $E_R$ is an $SU_L(2)$ singlet with 
the quantum numbers of the right-handed electron \cite{cliff}. Unification of 
gauge coupling constants then occurs at about $ 10^{11}$GeV with the 
standard normalization $g_1^2/g_2^2=3/5$ of the standard model $U(1)_Y$.
In the second scenario ($I_b$), the additional matter is taken to be 
$6 L+3 D_R$ vector-like representations, where $D_R$ is an $SU_L(2)$ singlet 
and $SU_c(3)$ triplet with the quantum numbers of the right handed 
$d$ quark. In this case, unification of gauge coupling constants occurs 
at about $10^{12}$GeV with the unconventional normalization of the standard 
model $U(1)$ $g_1^2/g_2^2=3/11$ . Scenario $I_b$ is inspired by an explicit 
$Z_3$ orientifold model \cite{FERNANDO} with this 
latter property, though the model does 
not have all the properties discussed in the next paragraph.

The soft supersymmetry-breaking terms for type I theories
are known where the observable sector gauge group and all observable 
sector matter are associated with 9-branes, 5-branes or open strings 
linking 9-branes to 5-branes, and all 5-branes sit at the orbifold 
fixed point at the origin, so that duality transformations can be 
exploited to the full. We shall consider the case where there are 
only $5_i$-branes for one value of $i$, say $5_3$-branes, where 
$i$ labels the complex compact dimension wrapped 
by the 5-brane, and where there is a single 
overall modulus $T$. We shall also assume that the observable gauge 
group is entirely in the 9-brane sector 
and the cosmological constant $V_0$ is zero, so 
that $C=1$ in the notation of Brignole et al \cite{IBA:Spain}, and that 
the CP violating phases $\alpha_S,\alpha_T$ are zero. 
Then, the soft supersymmetry-breaking 
terms are universal and the same as in the large $T$ limit of the Calabi-Yau 
compactification of the weakly coupled heterotic string

\begin{equation}
M_{1/2}=\sqrt{3} m_{3/2} \sin \theta
\end{equation}
\begin{equation}
m^2_{0}=m^2_{3/2} \sin^2 \theta 
\end{equation}
\begin{equation}
A=-\sqrt{3} m_{3/2} \sin \theta
\end{equation}
where $M_{1/2},m_0$ and $A$ are the observable sector gaugino mass, 
scalar mass and trilinear scalar coupling, respectively, and $m_{3/2}$ 
is the gravitino mass. The Goldstino angle $\theta$ has been introduced  
by parametrizing the auxiliary fields $F^S$ and $F^T$  for the 
dilaton $S$ and modulus $T$ in the form
\begin{eqnarray}
F^S&=&\sqrt{3} m_{3/2} (S+\bar{S})\sin\theta , \nonumber \\
F^i&=&\sqrt{3} m_{3/2} (T+\bar{T})\cos\theta 
\end{eqnarray}
The effects of twisted sector moduli entering the gauge kinetic 
function and mixing with the $T$ modulus through a Green-Schwarz term have 
been neglected.

As mentioned earlier, a novel sparticle spectrum can arise when the 
renormalization group equations are run from an intermediate string scale 
of order $10^{11}$ or $10^{12}$GeV instead of $10^{16}$GeV, especially 
when unification at the intermediate scale is achieved by the introduction 
of extra matter states, even though the soft supersymmetry-breaking terms 
at the string scale are not novel. We shall present results for the 
dilaton dominated case $\theta=\frac{\pi}{2}$ and for a ``typical'' case 
$\theta=\frac{\pi}{4}$ but not for $\theta=0$, in which case the loop 
corrections become important.

Our parameters are the goldstino angle $\theta$, sign $\mu$ (which is 
not determined by the radiative electroweak symmetry-breaking constraint) 
\cite{Tam:Rad} where $\mu$ is the Higgs mixing parameter 
in the low energy superpotential 
and $\tan\beta$, the ratio of Higgs expectation values $\frac{<H_2^0>}{
<H_1^0>}$, if we leave $B$, the 
coefficient of the soft bilinear term associated with the Higgs mixing term,
 to be a free parameter to be determined by the 
minimization of the Higgs potential. Using (1),(2) and (3) as boundary 
conditions, the renormalization group equations are run 
from the unification scale to the electroweak 
scale and the sparticle spectrum is determined consistently with the 
constraints of correct electroweak symmetry breaking and experimental 
constraints on sparticle masses from unsuccessful searches at accelerators. 
The empirical lower bounds we use are $m_{\chi_1^{\pm}}>84$GeV, $
m_{\chi_1^0}>31.6$GeV, $m_{h_0}>89.3$GeV and $m_{\tilde{g}}>300$GeV \cite{
IBANEZ}.
Electroweak symmetry breaking is characterized by the extrema equations 

\begin{eqnarray}
\frac{1}{2}M_Z^2&=&\frac{\bar{m}^2_{H_1}-\bar{m}^2_{H_2}\tan^2 \beta}
{\tan^2 \beta -1}-\mu^2 \nonumber \\
-B\mu&=&\frac{1}{2}(\bar{m}^2_{H_1}+\bar{m}^{2}_{H_2}+2\mu^2)\sin 2\beta
\end{eqnarray}
where 
\begin{equation}
\bar{m}^2_{H_1,H_2}\equiv m^2_{H_1,H_2}+\frac{\partial \Delta V}{
\partial {v^2_{1,2}}}
\end{equation}
and $\Delta V=(64 \pi^2)^{-1} {\rm {STr}} M^4[ln (M^2/Q^2)-\frac{3}{2}]$ 
is the 
one loop contribution to the Higgs effective potential. Contributions 
from the third generation of particles and sparticles
are included.
The chargino mass matrix is 

\begin{equation}
M_{ch}=\left(\begin{array}{cc}\\
M_2 & \sqrt{2} m_W \sin\beta \\
m_W \cos\beta & -\mu 
\end{array}\right)
\end{equation}
and the neutralino mass matrix is 

$$\left(\begin{array}{cccc} \\
M_1 & 0 & -M_Z A_{11} & M_Z A_{21} \\
0  & M_2 & M_Z A_{12} &-M_Z A_{22} \\
-M_Z A_{11} & M_Z A_{12} & 0 & \mu \\
M_Z A_{21} & -M_Z A_{22} & \mu & 0
\end{array}\right)$$ 
with 
$$\left(\begin{array}{cc} \\
 A_{11} & A_{12} \\
A_{21} & A_{22}\end{array}\right)= 
\left(\begin{array}{cc} \\
\sin\theta_{W} \cos\beta & \cos \theta_W \cos\beta \\
\sin\theta_W \sin\beta & \cos\theta_W \sin\beta 
\end{array}\right)$$
also, the stau mass matrix is given by
\begin{equation}
{\cal M}_{\tau}^2=\left(\begin{array}{cc} \\
{\cal M}_{11}^2 & m_{\tilde{\tau}}(A_{\tau}+\mu \tan\beta) \\
m_{\tilde{\tau}}(A_{\tau}+\mu \tan\beta) & {\cal M}_{22}^2
\end{array}\right)
\end{equation}
where ${\cal M}^2_{11}=m^2_L+
m^2_{\tau}-
\frac{1}{2}(2M_W^2-M_Z^2)\cos 2 \beta$ and ${\cal M}^2_{22}=
m^2_{e_R}+m^2_{\tau}+(M_W^2-M_Z^2)\cos2\beta$.

We now discuss the qualitative features of the sparticle spectrum 
for the two scenarios, type $I_a$ and $I_b$, described earlier 
and compare with the 
eleven dimensional supergravity limit of M-theory corresponding to the 
strongly-coupled heterotic string and with the weakly-coupled 
heterotic string 
in the large $T$ limit of Calabi-Yau compactification.

\begin{figure}
\epsfxsize=6in
\epsfysize=6.5in
\epsffile{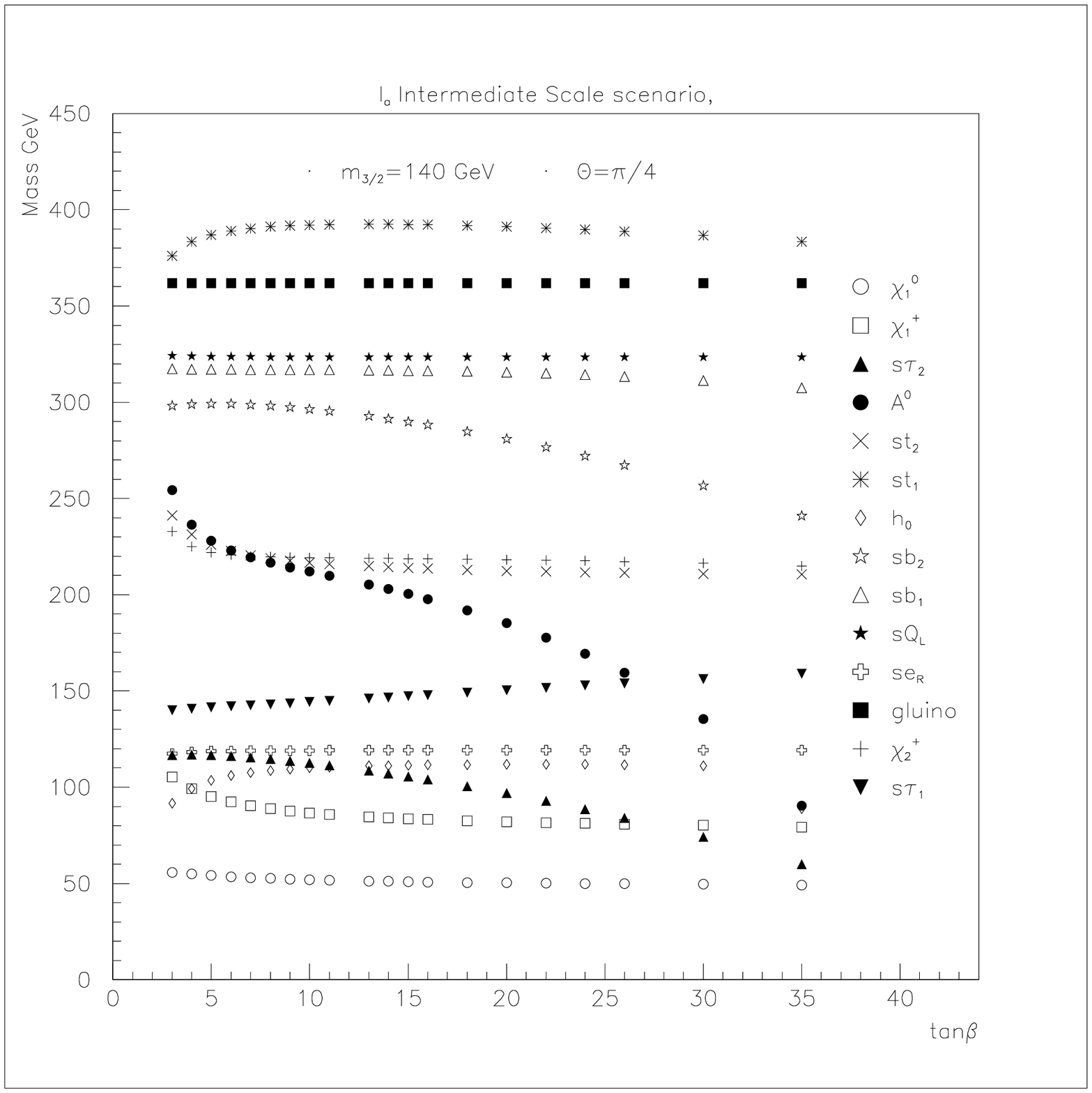}
\caption{$I_a$ scenario sparticle spectrum vs $\tan\beta$ for 
$m_{3/2}=140GeV,\mu>0,\theta=\pi/4$}
\label{Iapi4}
\end{figure}

\begin{figure}
\epsfxsize=6in
\epsfysize=6.5in
\epsffile{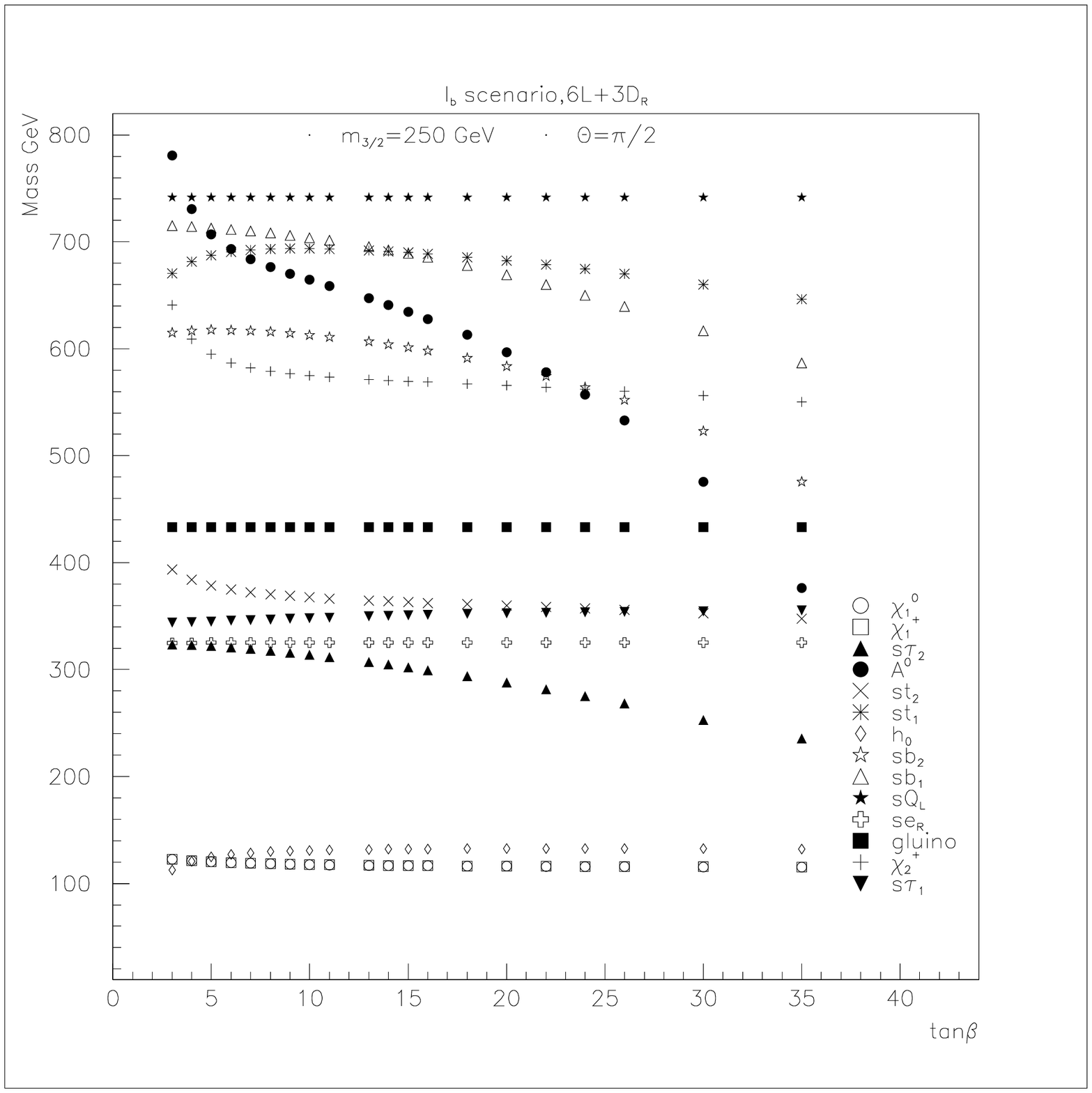}
\caption{Sparticle spectrum vs $\tan\beta$ $I_b$ scenario with 
$6L+3D_R$, $m_{3/2}=250GeV,\mu>0,\theta=\pi/2$}
\label{Ibpi2}
\end{figure}

\begin{figure}
\epsfxsize=6in
\epsfysize=6.5in
\epsffile{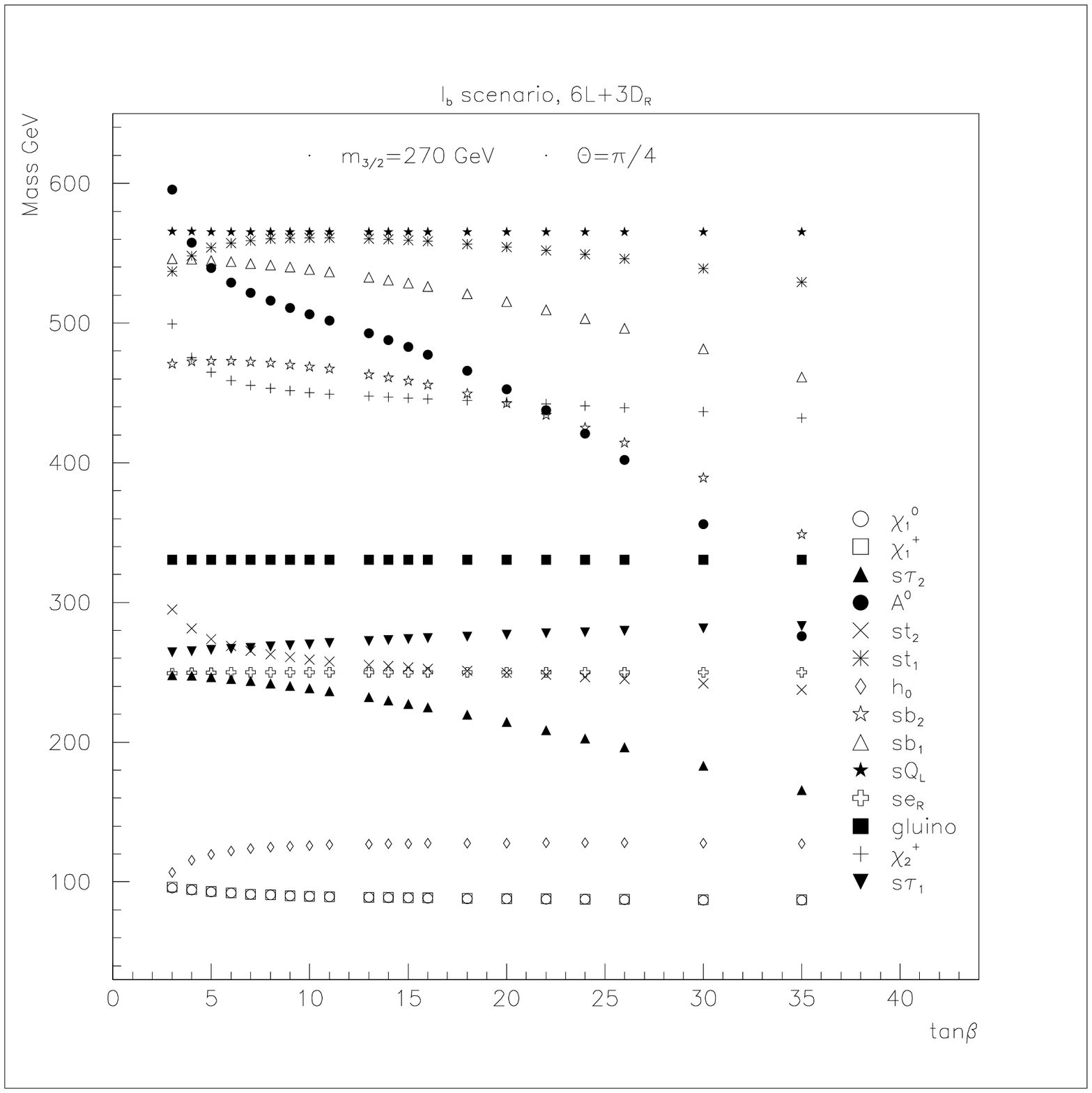}
\caption{Sparticle spectrum vs $\tan\beta$ $I_b$ scenario with 
$6L+3D_R$, $m_{3/2}=270GeV,\mu>0,\theta=\pi/4$}
\label{Ibpi4}
\end{figure}

The results of our calculation of the sparticle spectra arising from the 
different scenarios are 
presented in Figs.\ref{Iapi4}, \ref{Ibpi2}, \ref{Ibpi4}. 
For the $I_a$ scenario the sparticle spectra for $\theta=\pi/4$ and 
$\theta=\pi/2$ differ very little apart 
from the overall scale which is determined by the 
gravitino mass. We therefore only present the sparticle spectrum 
for $\theta=\pi/4$ in $I_a$ scenario.
We find also that the choice of sign of 
$\mu$ makes little difference to the spectra. Qualitatively, the 
principal features are as follows:

\begin{itemize}
\item The CP odd Higgs $(A^0)$ is much lighter in the $I_a$ scenario 
than the $I_b$. For the most part it is lighter than the lightest 
stop $(st_2)$ in the $I_a$ scenario, whereas in $I_b$ it is much 
heavier than $(st_2)$.

\item The lightest stau $(s\tau_2)$ is also lighter in the $I_a$ 
scenario than in $I_b$. In the former it is closer to the lightest 
chargino $(\chi_1^+)$, whereas in the latter it is much heavier than 
$(\chi_1^+)$.

\item Moreover, in $I_a$ the $(\chi_1^+)$ is much heavier than the 
lightest neutralino $(\chi_1^0)$, whereas in $I_b$ they are almost 
degenerate, with $(\chi_1^0)$ being a few GeV lighter. This has 
important consequences for dark matter due to coannihilation effects.

\item In $I_a$ the gluino is 
almost the heaviest 
sparticle, whereas in $I_b$ it is in the middle of the spectrum 
(and lighter than $A_0$ for $\tan\beta<32$).

\end{itemize}

\begin{figure}
\epsfxsize=6in
\epsfysize=6.5in
\epsffile{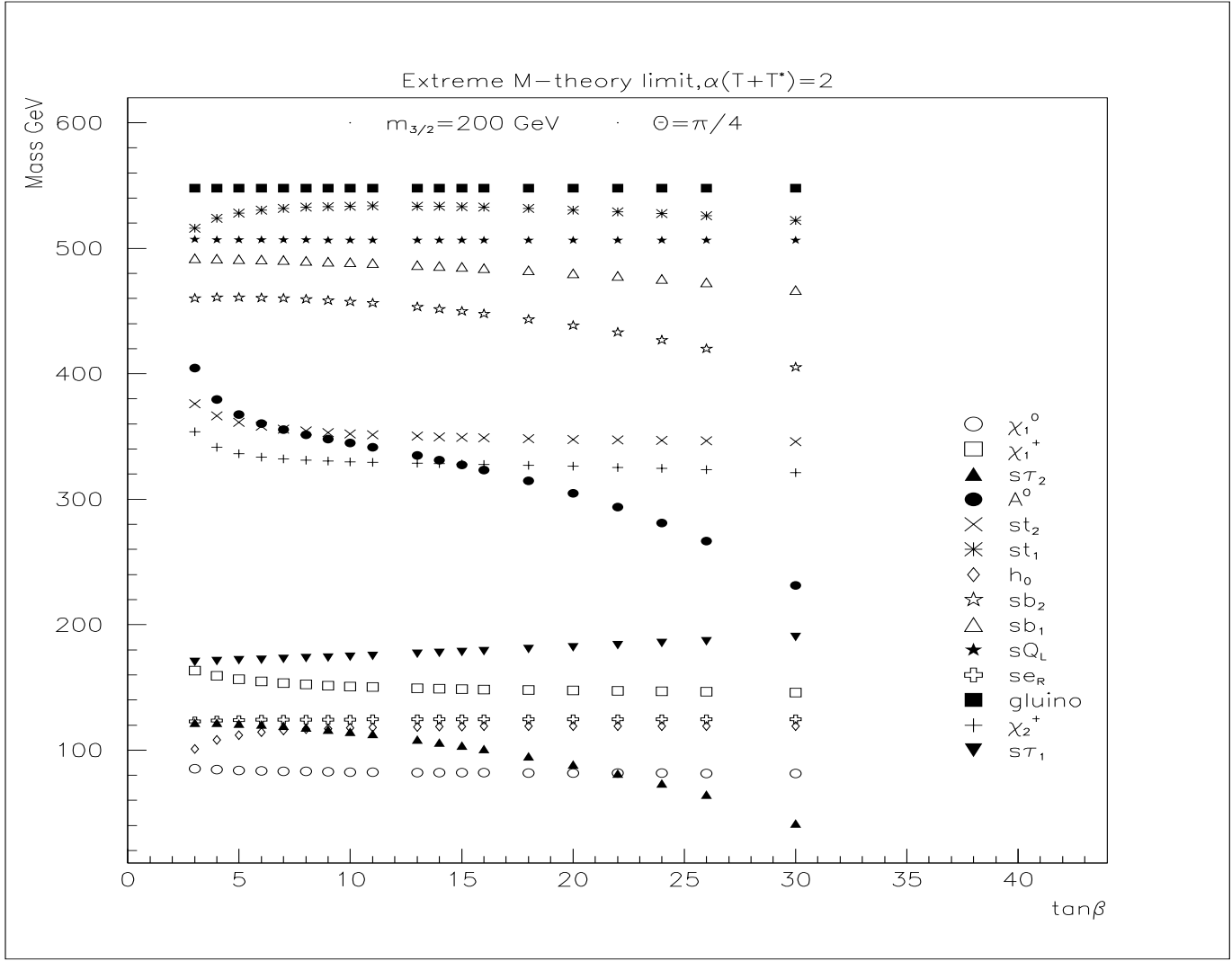}
\caption{Sparticle spectrum vs $\tan\beta$ in extreme 
M-theory limit, i.e $\alpha(T+\bar{T})=2$, $\theta=\frac{\pi}{4}$, 
$m_{3/2}=200GeV, \mu>0$.}
\label{extreme}
\end{figure}

\begin{figure}
\epsfxsize=6in
\epsfysize=6.5in
\epsffile{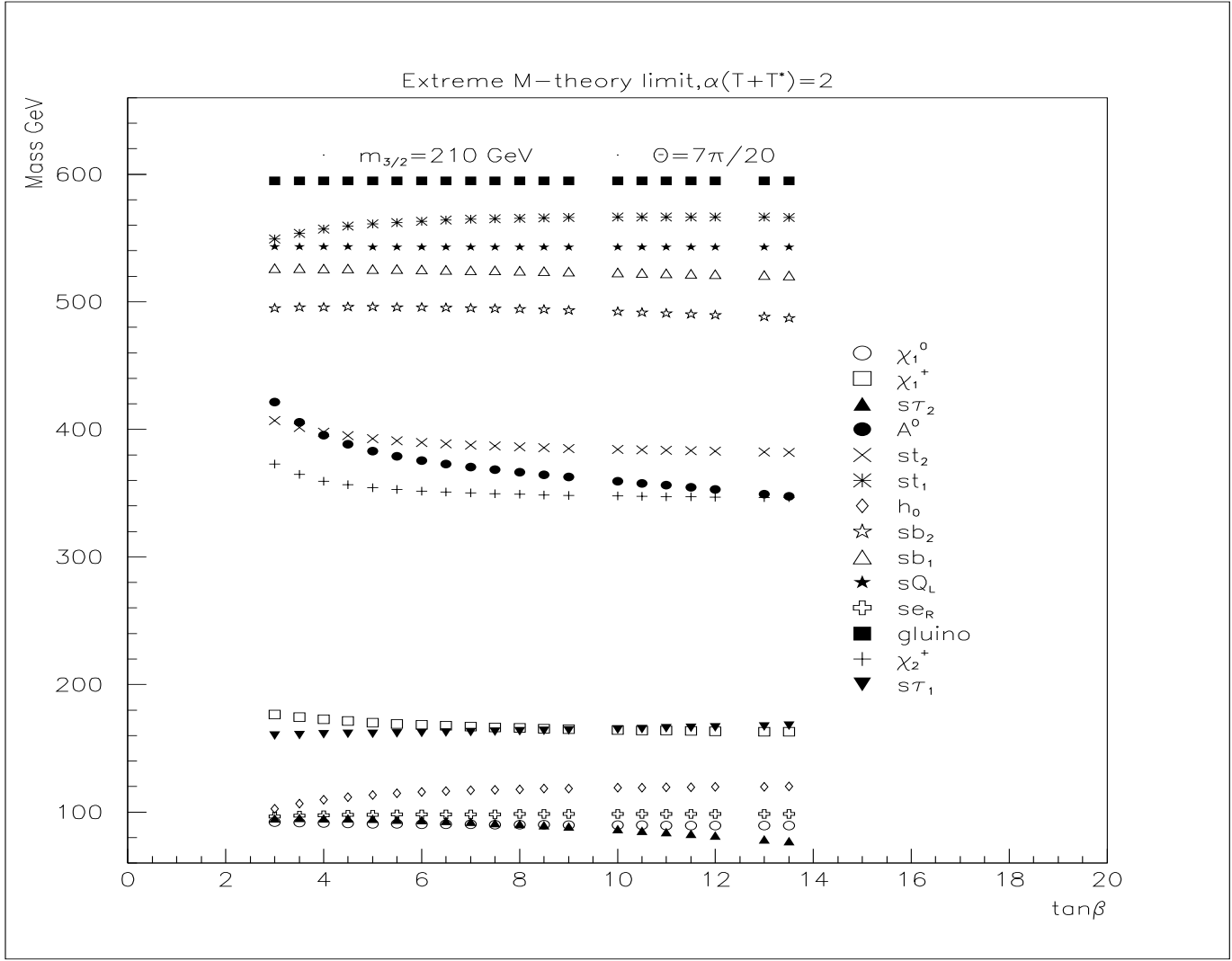}
\caption{Sparticle spectrum vs $\tan\beta$ in extreme 
M-theory limit, i.e $\alpha(T+\bar{T})=2$, $m_{3/2}=210GeV,\;\theta=
\frac{7\pi}{20},\;\mu>0$.}
\label{extremal}
\end{figure}

\begin{figure}
\epsfxsize=6in
\epsfysize=6.5in
\epsffile{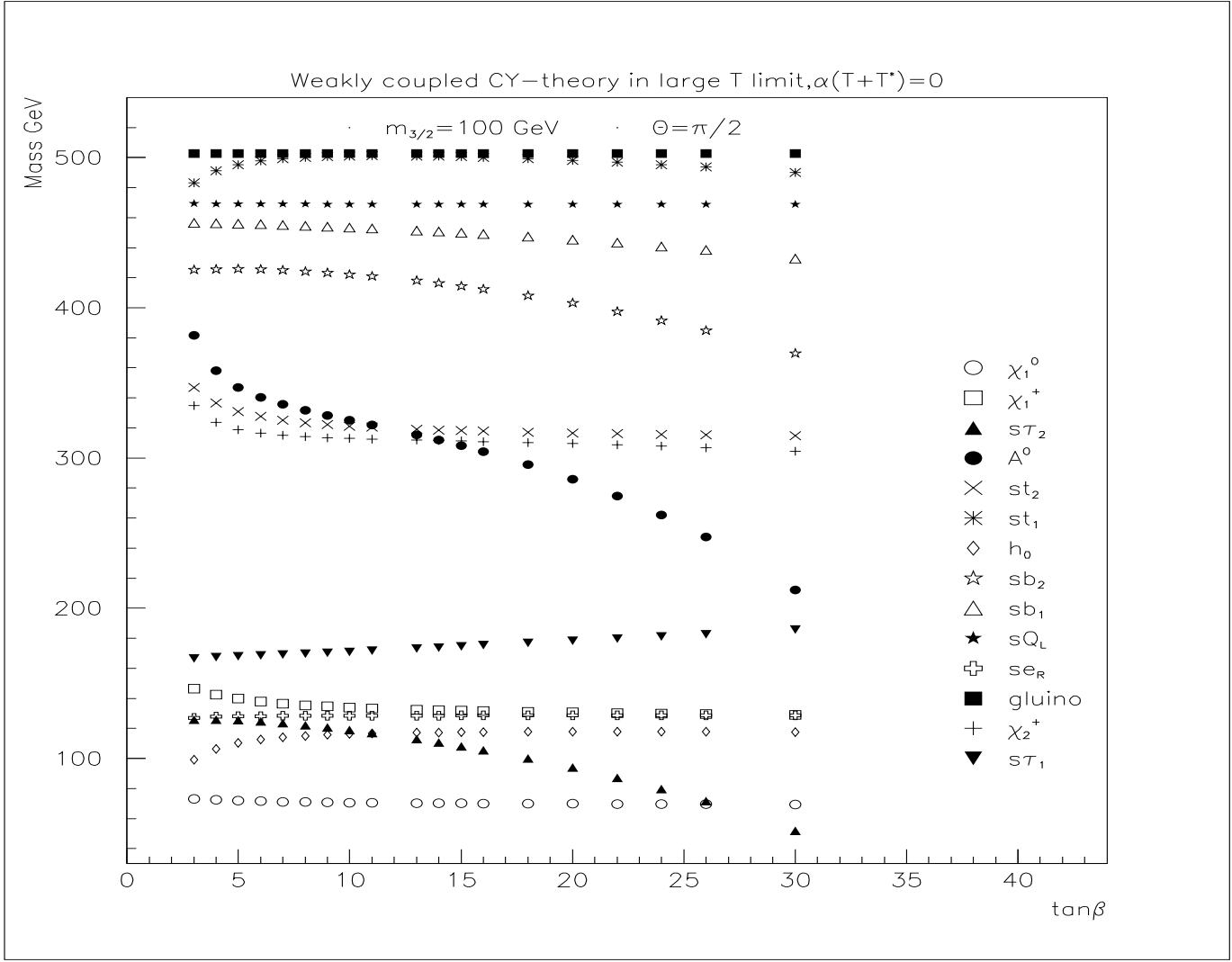}
\caption{Sparticle spectrum vs $\tan\beta$, $m_{3/2}=100GeV,\mu>0$, 
in the large T-limit of weakly-coupled CY space, $\theta=\pi/2$}
\label{WEAK1}
\end{figure}

\begin{figure}
\epsfxsize=6in
\epsfysize=6.5in
\epsffile{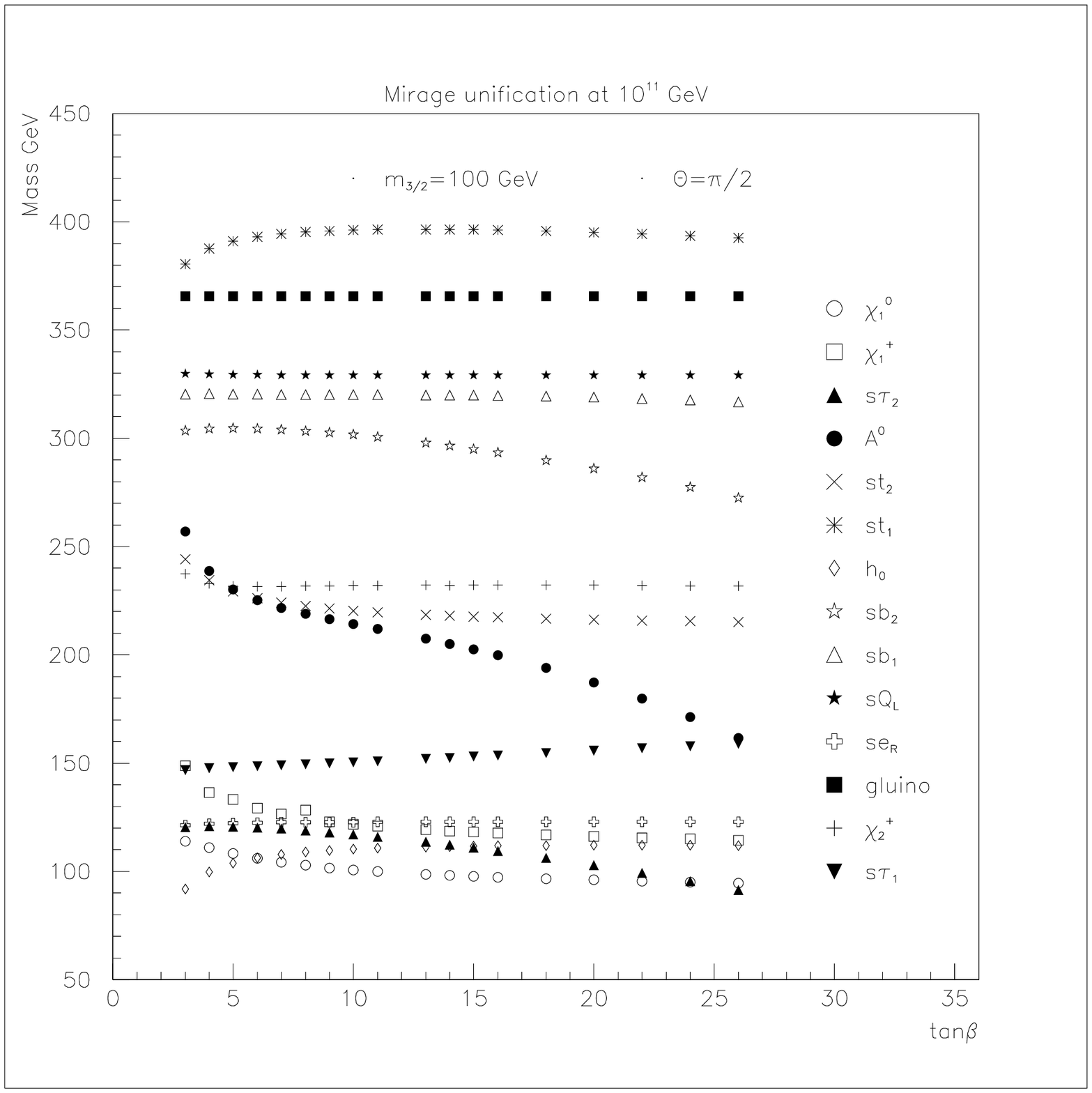}
\caption{Sparticle spectrum vs $\tan\beta$, in mirage unification scenario,
 $m_{3/2}=100GeV,\mu>0$.}
\label{MIR1}
\end{figure}

\begin{figure}
\epsfxsize=6.0in
\epsfysize=6.5in
\epsffile{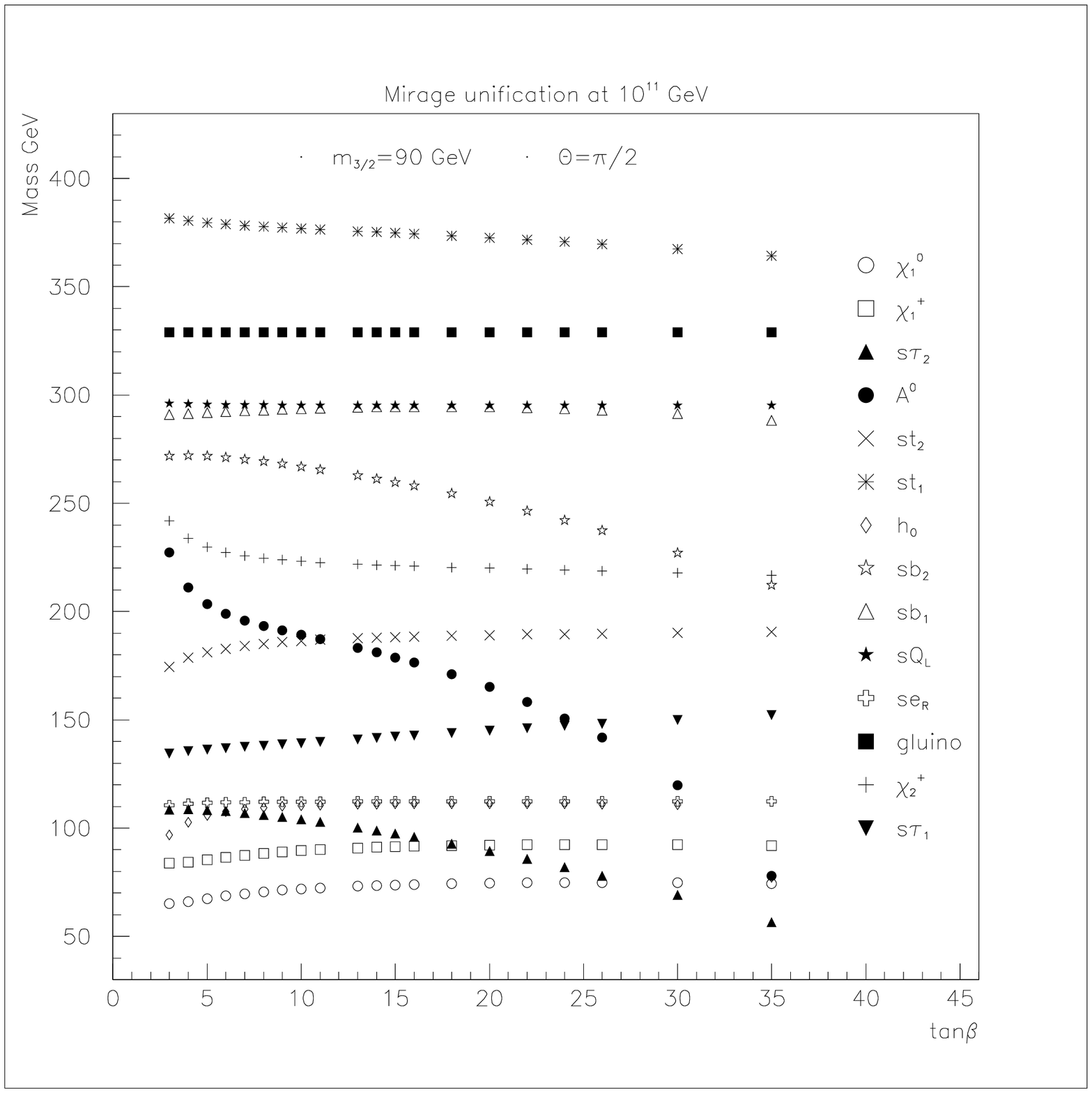}
\caption{Sparticle spectrum vs $\tan\beta$ in mirage unification 
scenario, $m_{3/2}=90GeV, \mu<0$.}
\label{MIR2}
\end{figure}

It is of interest to compare these spectra with those arising 
in other string scenarios. To allow this comparison we present 
the sparticle spectra for the extreme $M$-theory limit in Figs. \ref{extreme},
\ref{extremal},
for the large-T limit of weakly coupled string theory compactified 
on a Calabi-Yau space Figs. \ref{WEAK1}, and for the case 
of mirage 
unification \cite{LUIS}, 
with soft supersymmetry-breaking terms 
running from $10^{11}$GeV, in Figures \ref{MIR1},
\ref{MIR2} (In the case of mirage unification 
there is no extra matter, so the gauge couplings are $not$ unified 
at the string scale; the ``mirage'' of unification at $10^{16}$GeV 
is given by string loop effects.). In the extreme $M$-theory limit 
case the Goldstino mixing angle $\theta=\frac{\pi}{2}$ is not accessible 
without the scalar mass squared ($m_0^2$) becoming negative at the 
unification scale, hence breaking the electroweak gauge symmetry in 
models with the standard model gauge group.

The noteworthy qualitative features of this comparison are as follows:

\begin{itemize}

\item The $I_a$ spectrum is similar in most respects to that of the 
extreme $M$-theory case with $\theta=\frac{\pi}{4}$. 
However, in the $I_a$ case $s \tau_2$ is heavier than $\chi_1^+$ for 
$\tan\beta\leq 25$, whereas in the extreme $M$-theory case it is 
always lighter than $\chi_1^+$. Furthermore, in the $M$-theory case,
for $\tan\beta>23$, $(s\tau_2)$ becomes the lightest supersymmetric 
particle (LSP).

\item Most of the foregoing features are insensitive to the Goldstino 
angle. However, in the extreme 
M-theory case \cite{BKL} (see Fig.(\ref{extremal})) with 
$\theta=\frac{7\pi}{20}$ we have $m_0\ll M_{1/2}$, the common gaugino mass  
, whereas for $\theta=\frac{\pi}{4}$ we have 
$m_0\sim M_{1/2}$ and this difference 
produces some qualitative changes. For example, $s\tau_2$ becomes the 
LSP  for $\tan\beta>9$ when $\theta=\frac{7\pi}{20}$. Also $A^0$ is 
now heavier than when $\theta=\frac{\pi}{4}$, and $st_2$ is lighter 
than $A^0$, as is the case in the $I_b$ scenario.

\item The spectra deriving from the extreme $M$-theory limit with 
$\theta=\pi/4$ (Fig. \ref{extreme}) are similar in most respects  to those 
deriving from the weakly coupled case with $m_{3/2}=100$GeV, 
$\theta=\frac{\pi}{2}$ shown in Fig. \ref{WEAK1}.

\item The spectra arising in the mirage unification scenario  
(with $\mu>0$) are similar in most respects to those in $I_a$. 
However, for mirage unification $s\tau_2$ is lighter than 
$\chi_1^+$, whereas it is heavier than $\chi_1^+$ in $I_a$ for 
$\tan\beta<25$. 

\item The spectra arising in the two mirage unification scenarios, 
$\mu>0$ and $\mu<0$, are similar in most respects. However, for 
$\mu>0$ $\chi_1^+$ is always heavier than $s\tau_2$, whereas for 
$\mu<0$ $\chi_1^+$ is lighter than $s\tau_2$ for $\tan\beta<17$. 
Also, for small $\tan\beta$ the masses of $A^0$ and $st_2$ are very 
similar for $\mu>0$, but very different for $\mu<0$.


\end{itemize}

Assuming $R$-parity conservation 
the LSP is stable,  and consequently if it is neutral 
can provide a good dark matter candidate. 
We assume that the dark matter is in the form of neutralinos. 
The lightest neutralino  is a linear combination of the 
superpartners of the 
photon, $Z^0$ and neutral-Higgs bosons,
\begin{equation}
\chi_1^0=N_{11} \tilde{B}+N_{12}\tilde{W}^3+N_{13}\tilde{H}_1^0+N_{14}\tilde{H}_2^0
\end{equation}

For both $I_a$ and $I_b$ scenarios 
the lightest neutralino is the LSP for most of the parameter space
and for the mirage unification $\chi_1^0$ is not the LSP only for 
$\tan\beta>25.$ 
For these cases one can calculate the resulting 
relic abundance.

When the observational data on temperature fluctuations, 
type Ia supernovae, and gravitational lensing are combined with 
popular cosmological models, the dark matter relic abundance ($\Omega_{LSP}$) 
typically satisfies \cite{PERL}
\begin{equation}
0.1\leq \Omega_{LSP} h^2 \leq 0.4
\label{COSMO}
\end{equation}

We calculated the relic abundance of the lightest neutralino 
in the scenarios we have considered using standard 
techniques \cite{MARK}.
When these results are confronted with the (model-dependent) 
bounds (\ref{COSMO}) derived from the observational data 
further constraints on the parameters $m_{3/2},\tan\beta,\mu,\theta$ are 
obtained and these give new constraints on the sparticle spectrum.

\begin{figure}
\epsfxsize=6.5in
\epsfysize=7.5in
\epsffile{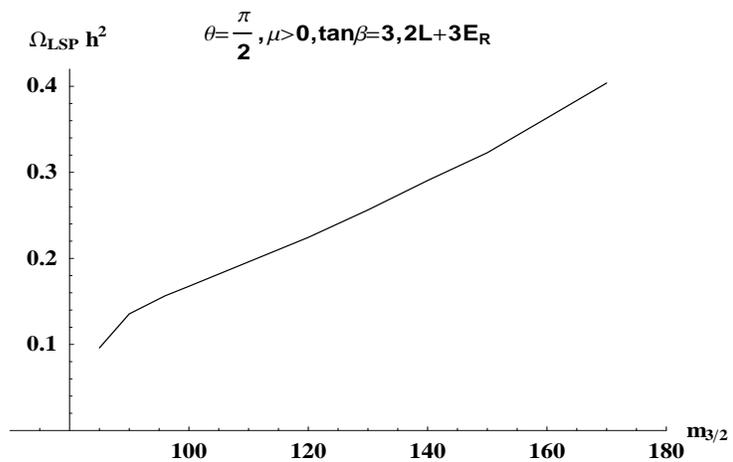}
\caption{Relic abundance of LSP vs $m_{3/2}$, $I_a$ scenario with 
$2L+3E_R$, $\tan\beta=3, \mu>0, \theta=\pi/2$}
\label{relic1}
\end{figure}

Let us start with the $I_a$ scenario. 
We first present the results of a calculation of the relic 
abundance for the lightest neutralino as a function of the 
gravitino mass $m_{3/2}$ for two representative values of 
$\tan\beta$, $\tan\beta=3$ and $\tan\beta=10$.
In Fig.\ref{relic1} we plot the relic 
abundance  for the lightest neutralino
versus the gravitino mass, $m_{3/2}$, for $\theta=\pi/2$ and 
$\tan\beta=3,\mu>0$. 
The upper and lower limit (\ref{COSMO}) 
on the relic abundance constrain $m_{3/2}$ to lie 
in the interval $85\leq m_{3/2}\leq 170$GeV see Table 1.. 

\begin{table}
\begin{center}
\begin{tabular}{|c||c|c|c|}  \hline\hline
Mass &{\bf $\theta=\frac{\pi}{2},\tan\beta=10,\mu<0$} & {\bf $\theta=\frac{\pi}{2},\tan\beta=10,\mu>0$} & {\bf $\theta=\frac{\pi}{2},\tan\beta=3,\mu>0$} \\
\hline\hline
$m_{3/2}$& 113GeV-170GeV  & 100GeV-167GeV & 85GeV-170GeV \\
\hline\hline
$m_{\chi_1^0}$& 54GeV-87.5GeV & 52.5GeV-89.1GeV& 49GeV-93.5GeV \\
\hline\hline
$m_{\chi_1^{\pm}}$& 87GeV-150GeV & 88GeV-158GeV &  93GeV-174GeV\\
\hline\hline
$m_{h^0}$& 115GeV-123GeV & 110GeV-121GeV & 89GeV-105GeV \\
\hline\hline
\end{tabular}
\end{center}
\caption{Bounds on  sparticle masses resulting from Eq. ($\ref{COSMO}$.)}
\end{table}

If instead we take 
 $\tan\beta=10$, $100{\rm GeV}\leq m_{3/2} \leq 167$GeV for $\mu>0$ 
whereas for $\mu<0$
 $113{\rm GeV}\leq m_{3/2} \leq 170$GeV.
In this case one obtains the bounds on the sparticle masses exhibited in 
Table 1.

\begin{figure}
\epsfxsize=6.5in
\epsfysize=7.5in
\epsffile{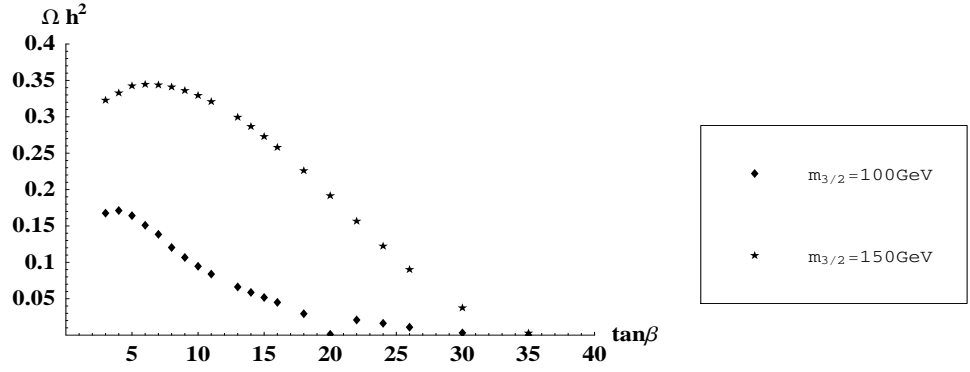}
\caption{Relic abundance of LSP vs $\tan\beta$, $I_a$ scenario with 
$2L+3E_R$, $m_{3/2}=100,150$GeV, $\mu>0, \theta=\pi/2$}
\label{relic2}
\end{figure}

The lightest stau for $\mu>0$ is in the range 
$114{\rm GeV}\leq m_{{\tilde{\tau}}_2}\leq 185$GeV. 
For the same value of $\tan\beta$ the total detection rates 
for a typical $^{73}$Ge 
detector are in the range $0.07 (2\times 
10^{-2})-4.8\times 10^{-3}(7\times 10^{-4})$events/Kg/Day
for $\mu<0(>0)$ respectively. The lightest supersymmetric particle is 
almost a Bino for both signs of $\mu$. For $\mu<0$ the Higgsino component 
is a little bit larger than for $\mu>0$ but still the LSP is essentially 
almost a Bino. 
In Fig. \ref{relic2} we plot the relic 
abundance versus $\tan\beta$ for fixed gravitino mass. From this 
figure  we see that 
for smaller values for the gravitino mass (i.e the lighter the spectrum) 
$\tan\beta$ is restricted to small values.

In the $I_b$  scenario and for $
\mu>0$ the constraints from (\ref{COSMO}) have dramatic consequences. 
As noted earlier
there is an almost exact degeneracy of the lightest chargino with the 
lightest neutralino, $M_{{\tilde{\chi}}_1^{\pm}}-M_{
{\tilde{\chi}}_1^{0}}\leq 3$GeV. Because of this 
coannihilation effects \cite{
YAMA} become important and the 
resulting relic abundance is very small. Therefore, if the lightest 
neutralino (which is almost 
Wino in this case) makes up most of the non-baryonic dark matter in 
the universe this model is 
excluded. 

In the mirage unification scenario (without extra matter) cosmological 
constraints prefer a low $\tan\beta$ and gravitino mass. For instance 
for $m_{3/2}=90$GeV, $\mu<0$ the relic abundance is in the range 
$0.12\geq \Omega_{LSP} h^2 \geq 0.01$ for $3\leq \tan\beta \leq 8$.
In this case, for $^{73}Ge,^{208}Pb,^{131}Xe$ detectors, 
detection rates of the neutralinos are in the range 
of order $10^{-1}-O(1)$events/Kg/day. This illustrates the fact that  
$\Omega_{LSP}h^2 \sim 
\frac{10^{-37}cm^2}{<\sigma_{anni} v>}$ and the neutralino annihilation 
cross section 
is roughly proportional to the neutralino scattering cross section. 
Thus as the LSP abundance decreases, its scattering cross section 
generally increases. 
For $\Omega_{LSP}h^2 \sim 0.1$ this results in an increased event rate.
Thus in this region of the parameter space even if the neutralino cosmic 
density is insufficient to close the universe, and other forms of dark matter
are needed, the prospects  of its direct detection 
in underground non-baryonic dark matter experiments could be enhanced.
This has also been noted by Gabrielli et al \cite{BOTT}.

This comparatively large direct detection rate is a consequence of 
the Higgsino component of the lightest neutralino being comparable to 
or even larger than the gaugino component.
As a result  the scalar cross section for the scattering of a 
neutralino with 
a nucleon through Higgs exchange increases. The scalar nucleon-LSP 
cross section is given by \cite{dbgvkal,BOTT}
\begin{equation}
\sigma_{scalar}^{(nucleon)}=\frac{8G_F^2}{\pi}M^2_W m^2_{red}\Biggl[
\frac{G_1(h_0)I_{h_0}}{m^2_{h_0}}+\frac{G_2(H)I_H}{m^2_H}+\cdots\Biggr]^2
\label{scalar}
\end{equation}
where 
\begin{eqnarray}
G_1(h_0)=(-N_{11}\tan\theta_W+N_{21})(N_{31}\sin\alpha+N_{41}\cos\alpha) \nonumber \\
G_2(H)=(-N_{11}\tan\theta_W+N_{21})(-N_{31}\cos\alpha+N_{41}\sin\alpha)
\end{eqnarray}
and 
\begin{equation}
I_{h_0,H}=\sum_{q}l_q^{h_0,H} m_q<N|\bar{q}q|N>
\end{equation}
and 
\begin{eqnarray}
l_q^{h_0}=\frac{\cos\alpha}{\sin\beta}\;\;\; 
l_q^H=\frac{\sin\alpha}{\sin\beta}\;\;{\rm for}\;\;\;\; q=u,c,t \nonumber \\
l_q^{h_0}=-\frac{\sin\alpha}{\cos\beta}\;\;\; 
l_q^H=\frac{\cos\alpha}{\cos\beta}\;\;{\rm for}\;\;\;\; q=d,s,b
\end{eqnarray}
In equation (\ref{scalar}) $m_{red}$ is the 
neutralino-nucleon reduced 
mass, $h_0,H$ denote the lightest Higgs and CP-even heavier Higgs 
respectively and $\alpha$ is the Higgs mixing angle.
We note also the $\tan\beta$ dependence
of the scalar neutralino-nucleon cross section 
$\sigma_{scalar}^{nucleon}$. For high values of $\tan\beta$ the 
corresponding cross section generically increases.
The ellipsis denotes
the contribution to the 
scalar cross-section through squark exchange which we have 
not written explicitly, although we included it in the calculations 
\cite{BKL}.
We note that, even if the scalar (spin-independent) 
interaction is the dominant one, the spin-dependent interaction through 
$Z$-exchange is also  
appreciable in this case since it is proportional to the difference
$[|N_{31}|^2-|N_{41}|]^2$ \cite{BKL}.

In summary, the three intermediate scale scenarios studied have sparticle 
spectra with striking qualitative features which distinguish them 
from each other and from the $M$-theory and weakly coupled heterotic 
string cases. Moreover, the composition of the lightest neutralino 
differs in the three scenarios. (It is almost Wino for 
the $I_b$ scenario, with 
a large Higgsino component for the mirage unification scenario, and almost 
Bino for the $I_a$ scenario, as well as for the $M$-theory and 
weakly-coupled heterotic string cases.) If we assume that the 
lightest neutralino provides the dark matter in the universe, constraints 
on the relic abundance put lower and upper bounds on the sparticle masses 
in each scenario. Also, the $I_b$ scenario is then excluded because 
coannihilation effects result in a too small relic abundance. 
Direct detection rates for the lightest neutralino in the $I_a$ scenario 
are similar to those for $M$-theory and weakly coupled heterotic string 
models. Interestingly, direct detection rates one or two orders of 
magnitude larger are obtained in the mirage unification scenario where the 
lightest neutralino has a large Higgsino component.

\section*{Acknowledgements}
This research is supported in part by PPARC.

\newpage


\newpage

\newpage

\newpage

\newpage

\newpage

\end{document}